\documentclass[prl,showpacs,floatfix,twocolumn]{revtex4}
\usepackage{graphicx}
\usepackage{dcolumn}
\usepackage{float}

\newcommand{\comment}[1]{}

\begin{document}

\title{\boldmath The role of magnetism in forming the $c$-axis spectral peak
at 400~cm$^{-1}$ in high temperature superconductors. \unboldmath}
\author{T.~Timusk}
\email{timusk@mcmaster.ca}
\affiliation{Department of Physics and Astronomy, McMaster University, Hamilton
ON Canada, L8S 4M1}
\author{C.C.~Homes}
\affiliation{Department of Physics, Brookhaven National Laboratory, Upton, NY
11973}
%

%
% The abstract goes here
%
\begin{abstract}
We discuss the peak at 400 cm$^{-1}$, which is seen in {\it c}-axis
conductivity spectra of underdoped high temperature superconductors. The model
of van der Marel and Munzar, where the peak is the result of a transverse
plasmon arising from a low frequency conductivity mode between the closely
spaced planes, fits our data well. Within the model we find that the
temperature dependence of the peak amplitude is controlled by in-plane
scattering processes. The temperature range where the mode can be seen
coincides with $T_s$, the spin gap temperature, which is lower than $T^\ast$,
the pseudogap temperature. As a function of temperature, the amplitude of the
mode tracks the amplitude of the 41 meV neutron resonance and the spin lattice
relaxation time, suggesting to us that the mode is controlled by magnetic
processes and not by superconducting fluctuations which have temperature scale
much closer to $T_c$, the superconducting transition temperature.
\end{abstract}

\pacs{74.25.Kc, 74.25.Gz, 74.72.-h}

\maketitle
%

%Outline:
\section{Introduction}

The first infrared measurements on ceramic samples of high temperature
superconductors early in 1987 showed two unexpected features. In addition to
the well known optical phonons of the perovskite lattice, two new features
could be seen. The first was a dramatic reflectance edge that appeared below
the superconducting transition temperature and the second, seen in many samples
with more than one copper-oxygen plane, was a peak in the 400 cm$^{-1}$ region,
much broader than a typical phonon. The peak frequency of the feature increases
with doping but we will refer to it as the 400 cm$^{-1}$ peak which is the
position in underdoped YBa$_2$Cu$_3$O$_{6.60}$. The reflectance edge was soon
explained as a plasma edge, resulting from the zero crossing of the real part
of the dielectric function where the positive contribution of the ionic
displacement current cancels the negative current due to the superconducting
condensate.\cite{bonn87prb} Since the super-current is a current of pairs, the
plasma edge has been termed the Josephson plasma
edge.\cite{tamasaku92,tachiki94} However, the appearance of the plasma edge is
a classical effect and it would be present in any conductor where the
scattering rate is sufficiently small, $1/\tau << \omega_p$.

The peak at 400 cm$^{-1}$ was ignored in the literature until the work of Homes
{\it et al.} \cite{homes93,homes95,homes95a,schutzmann95} where its properties
were described in some detail.  Recently, the two phenomena have been connected
by suggestions that the 400 cm$^{-1}$ peak is a {\it transverse} plasma
resonance associated with a coupling between the closely spaced planes of the
bilayers by van der Marel and by Munzar {\it et al.}
\cite{vandermarel_private,gruninger00,munzar99} A transverse plasma resonance
occurs in situations where two or more plasmas are geometrically separated.
This is in contrast to a multi-component plasma, for example one of light and
heavy holes in a semiconductor, which gives rise to a single longitudinal mode
with a frequency that depends on the two masses.  On the other hand, if the
light and heavy hole plasmas phase-separate into droplets of a heavy hole
plasma within a light hole medium two longitudinal and one transverse plasma
resonance will be seen.\cite{kirczenow79,zarate82}

Shibata and Yamada have reported the presence of two longitudinal Josephson
plasma resonances in SmLa$_{1-x}$Sr$_x$CuO$_{4-\delta}$ through the observation
of sphere resonances.\cite{shibata98} Powdered crystals, suspended in an
optically transparent matrix, show a resonance absorption at
$\omega=\omega_p\sqrt(3)$ where $\omega_p$ is the Josephson plasma
frequency.\cite{noh90,shibata98}

The material Shibata and Yamada used  to obtain the two separated plasmas has
single copper-oxygen planes separated by alternating Sm$_2$O$_2$ and
(La,Sr)$_2$O$_{2-\delta}$ blocking layers. When the material becomes
superconducting two resonances are seen. Both appear at $T_c$ and follow
mean-field temperature dependencies although the two curves are not identical
as shown in the inset of Fig.~2 of Ref.~\onlinecite{shibata98}.  The authors
attribute the resonances to the two different series Josephson junctions in the
two different blocking layers. The transverse plasma resonance in this material
has also been seen.\cite{shibata01,kakeshita01,dulic01}

Kojima {\it et al.} find that a 7~T magnetic field, applied along the {\it
c}-axis, has the effect of introducing a new mode at low frequency which is
interpreted in terms of transverse plasma resonance modulated by
vortices.\cite{kojima02} The presence or absence of a vortex between the layers
gives rise to two kinds of Josephson junctions which in turn leads to a
transverse Josephson resonance as originally proposed by van der Marel and
Tsvetkov.\cite{vandermarel96}  Kojima {\it et al.} also find that the spectral
weight of the 400 cm$^{-1}$ mode is enhanced in a magnetic field by some 20\%
while at the same time the spectral weight associated with the superconducting
condensate is reduced. They suggest this is due to a transfer of spectral
weight from the low frequency modes to the 400~cm$^{-1}$, which they assume to
be a transverse Josephson mode.\cite{kojima02}

Figure 1 shows the reflectance and the {\it c}-axis optical conductivity of a
YBa$_2$Cu$_3$O$_{6.60}$ crystal with $T_c=58$~K at a series of temperatures in
the region of the spectral peak at 400~cm$^{-1}$ from the work of Homes {\it et
al.}\cite{homes95} The rapid buildup of the spectral peak can be seen clearly
along with the reduction in the strength of the phonon at 320 cm$^{-1}$ which
corresponds to the out-of-plane vibrations of the planar oxygens.  The strength
of the chain oxygen peak at 280~cm$^{-1}$ is only slightly affected by the
buildup of the spectral peak. The 400 cm$^{-1}$ peak can also be seen in the
spectra of YBa$_2$Cu$_4$O$_8$,\cite{basov94,hauff96}
Pb$_2$Sr$_2$(Y,Ca)Cu$_3$O$_8$,\cite{reedyk94} the three layer
Tl$_2$Ba$_2$CaCu$_3$O$_{10}$,\cite{zetterer90} and the ladder compound
Sr$_{14-x}$Ca$_x$Cu$_{24}$O$_{41}$.\cite{osafune99} A weak feature has been
reported in Bi$_2$Sr$_2$CaCu$_2$O$_8$,\cite{zelezny99,zelezny01} and in
Bi$_2$Sr$_2$Ca$_2$Cu$_3$O$_{10}$ as well.\cite{boris02} It is not seen in the
one-layer La$_{2-x}$Sr$_x$CuO$_4$.

%
% Figure 1
%
\begin{figure}[t]
\vspace*{-0.4cm}%
\centerline{\includegraphics[width=4.0in]{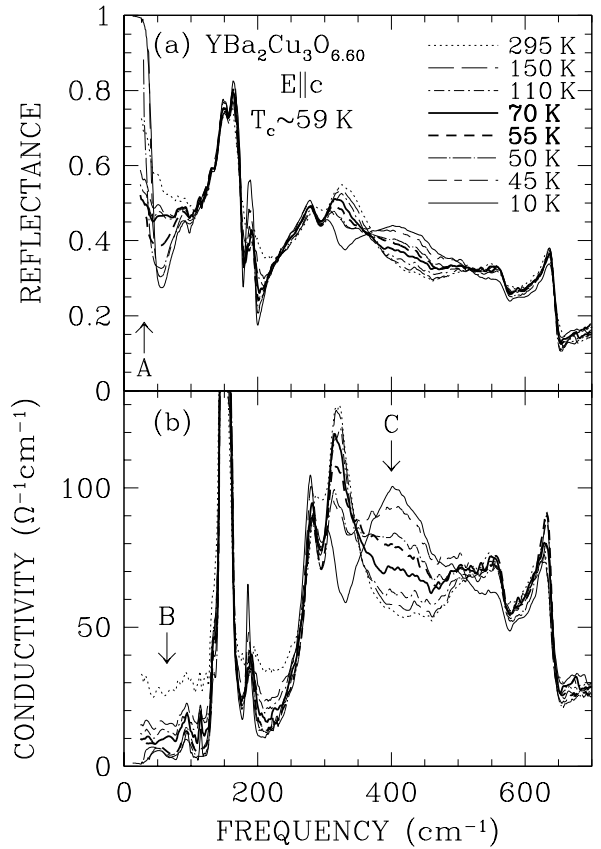}}%
\vspace*{-1.0cm}%
%\medskip
\caption{The reflectance (a) and the optical conductivity (b) at various
temperatures of an underdoped sample of YBa$_2$Cu$_3$O$_{6.60}$ for light
polarized along the {\it c} axis. The heavy solid line is just above the
superconducting transition temperature, the dashed line just below it.  Note
the appearance of the peak well above the superconducting transition
temperature which is 58~K for this sample. As the peak at 400~cm$^{-1}$ grows
the phonon at 320~cm$^{-1}$ weakens and shifts to lower frequencies.  We
interpret this in terms of changing local fields between the planes. The lower
frequency phonon at 280~cm$^{-1}$ is not affected by these fields. The peak at
400~cm$^{-1}$ cannot be seen above a temperature of 150~K. The arrow A is the
frequency used to estimate the condensate strength, the arrow B the pseudogap
depth and the arrow C the strength of the 400~cm$^{-1}$ peak.}%
\label{lumpfig1}
\end{figure}

An important contribution to the debate about the strange phonon anomalies and
the peak at 400 cm$^{-1}$ was made by van der Marel and
Tsvetkov\cite{vandermarel96} and by Munzar {\it et al.} \cite{munzar99} This
idea had two parts.  First, they assumed that the peak was due to a transverse
plasmon (intra-bilayer plasmon) resulting from the coupling between the two
planes of the bilayer system. The coupling across the blocking layers is weaker
and as a result of two unequal junctions in series, a charge imbalance was set
up in the unit cell which alters the local electric fields acting on the ions.
Model calculations showed that both the peak at 400~cm$^{-1}$ and the strengths
of nearby phonon lines were reproduced.\cite{munzar99} In particular, the
dramatic loss of intensity of the bond-bending mode of the planar oxygens at
320~cm$^{-1}$ was well accounted for.

The total spectral weight in the $250 - 700$~cm$^{-1}$ region is approximately
constant as the spectral peak develops. This means that it gains its spectral
weight from either the phonons in this spectral region or from a redistribution
of the electronic background. In Homes {\it et al.} it was assumed that the
phonons were the source whereas Bernhard {\it et al.} argued, based on a
different interpretation of the electronic background, that only 40\% of the
spectral weight is accounted for by phonons and the remainder from an assumed
lower electronic background.\cite{bernhard00a} As the raw reflectance and
ellipsometric data used in Bernhard {\it et al.} are nearly identical, the
different conclusions result from different assumptions as to the baseline in
the 400 cm$^{-1}$ region. These anomalies in the phonon spectral weight are
strictly a low temperature phenomenon.  It was shown by Timusk {\it et al.}
that at room temperature the strengths of the {\it c}-axis polarized phonons
had their expected magnitudes as determined by formal charges on the ions and a
shell model based on inelastic neutron spectroscopy.\cite{timusk95a}

The temperature dependence of the 400 cm$^{-1}$ mode has been discussed in the
literature by several authors. It was pointed out by Homes {\it et al.} that
while the mode is present above $T_c$ it appears at a temperature of 150~K,
which is much lower than that of the pseudogap and which can be seen at room
temperature.\cite{homes95a} Sch\"utzmann {\it et al.} suggested that the phonon
anomalies might be associated with the spin gap in the nuclear spin relaxation
rate $1/(T_1T)$ that occurs at 150~K.\cite{schutzmann95} Following up on this
idea, Hauff {\it et al.} showed that Zn substitution had a dramatic effect of
destroying the 400~cm$^{-1}$ peak while at the same time not changing the
pseudogap, which they defined as a low frequency depression of conductivity.
More recently, it has been suggested that the current $J_{bl}$ within the
bilayers that is responsible for the mode at 400 cm$^{-1}$ is a Josephson
current.\cite{munzar99,bernhard00a} It is difficult to see how such a current
can exist in the normal state up to 150~K. Note however that the existence of
the peak at 400 cm$^{-1}$ does not require that the intra-plane current
$J_{bl}$ is superconducting as pointed out by Gr\"uninger {\it et al.}
\cite{gruninger00}.  We will return to the problem of the interlayer current
below.

The focus of the remainder of the paper is on the temperature dependence of the
400~cm$^{-1}$ mode. First we emphasize that the mode appears at 150~K, well
above the superconducting transition temperature. This fact is difficult to
reconcile with the idea of a transverse Josephson plasmon unless some kind of
2D superconductivity exists above $T_c$ in the planes. We discuss the evidence
for this kind of superconductivity. Then we examine models within the van der
Marel/Munzar (vdMM) picture where the oscillator that represents the
intra-bilayer current $J_{bl}$ is allowed to have a width.  Next we focus on
the temperature dependence of the mode. We note, as first suggested by
Sch\"utzmann {\it et al.}, that the mode follows the temperature dependence the
NMR $1/(T_1T)$ connecting the mode intimately to the spin gap and magnetism but
we find that the mode intensity follows even more closely the intensity of the
41~meV neutron mode. In contrast, the Knight shift temperature scale is higher
and more closely related to the pseudogap in the conductivity. Finally, we
offer some speculations on what might cause this intimate connection between
intra-bilayer transport and magnetism.

\section{Results and discussion}
%
%\subsection{superconducting fluctuations}
%
The idea that the transverse plasmon giving rise to the 400~cm$^{-1}$ peak is
the result of the Josephson effect connecting the two superconducting planes
has some attractive features. The model of Munzar {\it et al.} shows that there
is a narrow  peak near zero frequency corresponding to a current $J_{bl}$
between the closely spaced layers. It first appears at 150~K and grows rapidly
at $T_c$. However this requires the existence of some kind of superconductivity
in the planes at temperatures as high as 150~K. Moreover this superconductivity
has to be nearly fully developed at $T_c$ with a plasma frequency of
1000~cm$^{-1}$ at $T_c$ rising to 1200~cm$^{-1}$ at 4~K.\cite{munzar99} An
obvious source for this above-$T_c$ superconductivity is superconducting
fluctuations. However there is little evidence for strong superconducting
fluctuations in the {\it ab}-plane properties over a broad frequency
range.\cite{salamon_review} The only work we are aware of is the paper by
Corson {\it et al.}, who report a weak effect in underdoped
Bi$_2$Sr$_2$CaCu$_2$O$_8$ with an onset temperature around 92~K (in a sample
with $T_c=74$~K) and a $\sigma_2$ magnitude of some 18\% of the low temperature
value measured in the terahertz region.\cite{corson99} A similar narrow range
of temperatures has been reported by Sugimoto {\it et al.} \cite{sugimoto02} in
STM measurements of magnetic flux and by Zhang {\it et al.} \cite{zhang01} with
measurements of the Nernst effect. While there is evidence for vorticies above
$T_c$ we are not aware of any transport measurements reporting {\it ab}-plane
superconductivity on the temperature scale of $T_s$.  Finally the presence of
the peak in the non-superconducting ladder compound
Sr$_{14-x}$Ca$_x$Cu$_{24}$O$_{41}$\cite{osafune99} speaks against a direct
superconducting origin of the peak.  Thus it seems unlikely that
superconductivity is the source of the 400~cm$^{-1}$ peak.

%\subsection{study of transport between the bilayers using Munzar's model}
%
% Figure 2
%
\begin{figure}[t]
\vspace*{2.5cm}%
\centerline{\includegraphics[width=3.6in]{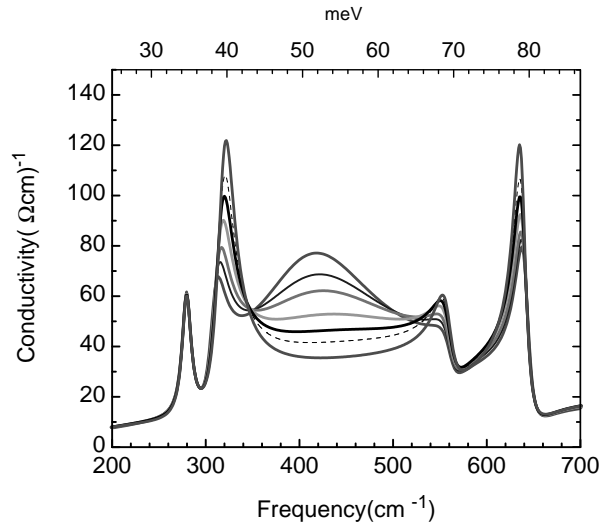}}%
\vspace*{-2.0cm}%
%\medskip
\caption{Model calculation of the influence of mode width on the peak at
400~cm$^{-1}$.  Starting from the parameters of Munzar {\it et al.}
(Ref.~\onlinecite{munzar99}), top curve, we have added a damping to the mode at
the origin.  The values for the damping are 25, 50, 100, 150, 200, and
300~cm$^{-1}$.  The mode weakens as the damping increases.}%
\label{lumpfig2}
\end{figure}

One of the advantages of the model of van der Marel and Munzar\cite{munzar99}
is that it allows a study of the transport between the layers in a double-layer
system, independent of the properties of the blocking layers. We will use the
model to study that transport by investigating the parameters that control the
properties of the transverse plasmon. In the original paper it was assumed that
the current between the layers was a Josephson current between two coupled
superconductors. The only parameter controlling the motion of charge between
the closely spaced planes in that picture was the Josephson plasma frequency
$\omega_{bl}$. Classically, the current was due to a Drude term with zero
damping parameter and a plasma frequency $\omega_{bl}$.

We will next examine the dynamic properties of the transverse plasmon that
gives rise to the 400 cm$^{-1}$ peak within the Munzar model. In the original
model the transverse plasmon was the result of a mode at the origin of zero
width and a plasma frequency of 1200 cm$^{-1}$. Using the parameters of
Ref.~\onlinecite{munzar99} we have recalculated the conductivity in the 200 to
700 cm$^{-1}$ region adding larger and larger amounts of width to the mode.
Fig. 2 shows the results of this calculation. Comparing this with Fig. 1 we see
that broadening the mode in the model matches the increasing temperature in the
experimental spectra. In other words the temperature dependence of the observed
transverse plasmon can be modeled by a low frequency Drude-like mode whose
width increases approximately linearly from zero to 300 cm$^{-1}$ and from low
temperature to 150 K. We have also examined the effect on the mode of a shift
away from zero frequency. Here the changes are much larger and the center
frequency of the mode has to be less than 50 cm$^{-1}$ to agree with the
experimental data.

As a reference we can look at the scattering rate associated with the in-plane
conductivity.\cite{ito93} With an in-plane plasma frequency of 7000
cm$^{-1}$\cite{ito93} we can estimate the scattering rates to vary from 25
cm$^{-1}$ at 62 K to 208 cm$^{-1}$ at 150 K. Comparing these figures with the
model shown in Fig. 2 we can predict the size of the 400 cm$^{-1}$ peak from
the model and the scattering rates calculated from dc transport. These
predicted values are shown as crosses in Fig. 3. While the agreement with the
observed peak size is rough, at least the calculation shows that in-plane
scattering is a possible source of the destruction of the 400 cm$^{-1}$ peak as
the temperature increases.
%
% Figure 3
%
\begin{figure}[t]
\vspace*{2.5cm}%
\centerline{\includegraphics[width=4.2in]{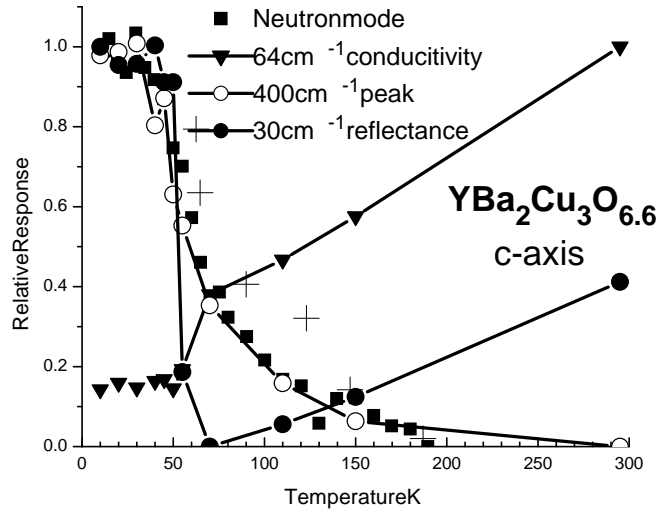}}%
\vspace*{-2.8cm}%
%\medskip
\caption{Response of various quantities as a function of temperature.  Three
temperature scales can be identified.  The largest is around 300~K and is
associated with the pseudogap which is approximated as the conductivity at
64~cm$^{-1}$.  The next governs the intensity of the 400~cm$^{-1}$ peak and
neutron mode. It has a value of approximately 150~K.  The lowest one is
associated with superconductivity and has a value of 60~K in this underdoped
sample. We use the 30~cm$^{-1}$ reflectance to approximate the superconducting
condensate density. The crosses show an estimate of the 400 cm$^{-1}$ peak
amplitude based on broadening by in-plane scattering.}%
\label{lumpfig3}
\end{figure}

We want to follow up the suggestion of Sch\"utzmann {\it et al.} and
investigate the role of magnetism in promoting the intra-bilayer currents. We
start by plotting in Fig.~3 several quantities that can be extracted from the
$c$-axis optical properties. The most important of these for the purposes of
the present paper is the amplitude of the 400~cm$^{-1}$ mode. We have plotted,
as a function of temperature, the conductivity at the maximum of the
400~cm$^{-1}$ peak, minus its room temperature value, normalized to its low
temperature value based on the 1993 reflectance measurements of Homes {\it et
al.}\cite{homes93} It can be seen that the peak first appears at 150~K, some 90
degrees above the superconducting transition, has its most rapid development
near $T_c$ where its amplitude is almost exactly 50\% of the saturation value,
which it reaches quickly below $T_c$.

We have also plotted in Fig.~3 a rough estimate of the superconducting
condensate density based on the reflectance at low frequency (30~cm$^{-1}$)
normalized to unity at 10 K and to zero at minimum reflectance at $T_c$. We
note that the condensate density rises rapidly at $T_c$ with a very narrow
pre-transition region of no more that 10~K. The slow rise of the low frequency
reflectance above $T_c$ is due to changes in the optical properties associated
with the pseudogap.

%\subsection{The relationship to magnetic phenomena}

%
% Figure 4
%
\begin{figure}[t]
\vspace*{2.5cm}%
\centerline{\includegraphics[width=4.2in]{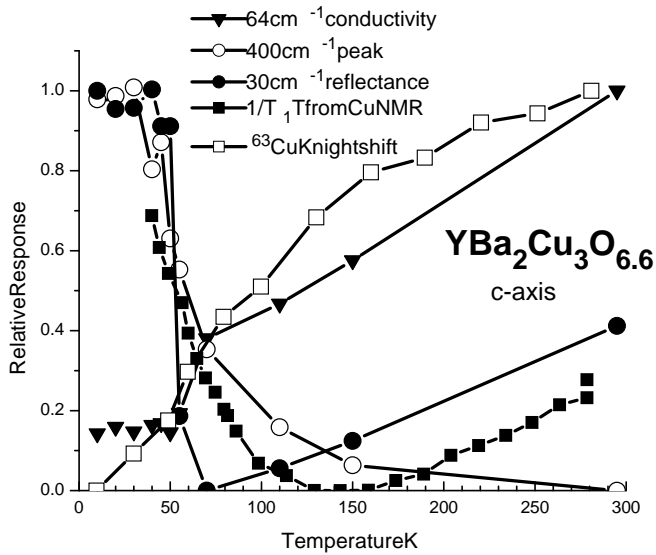}}%
\vspace*{-2.8cm}%
\caption{Two temperature scales seen in NMR.  The higher scale seen in the
Knight shift plotted as open squares and the pseudogap plotted as triangles and
approximated here as the 64~cm$^{-1}$ conductivity.  The Knight shift
depression starts at room temperature in this sample with an oxygen content of
6.60.  The relaxation time, shown as solid squares, is plotted as
$[1-1/(T_1T)]$.  It starts to deviate from the uniform high temperature form at
around 150~K and fits better to the 400~cm$^{-1}$ peak intensity than the
Knight shift.}%
\label{lumpfig4}
\end{figure}

Also plotted in Fig.~3 is the pseudogap amplitude defined as the conductivity
at 64~cm$^{-1}$ normalized to unity at room temperature.  The pseudogap
amplitude drops smoothly from room temperature with a rapid step in the
transition region to superconductivity.  This step may be due to the formation
of the superconducting gap as spectral weight is removed from the low frequency
region to the delta function peak at the origin.

We see from Fig.~3 that three temperature scales are involved in the
electrodynamics of {\it c}-axis transport. The highest of these describes the
pseudogap and has been traditionally called $T^\ast$.  At this doping level
$T^\ast$ is approximately 300~K. The scale associated with the 400~cm$^{-1}$
peak is lower and we will call it $T_s$ for reasons that will become clear
below.  The phenomena that follow the $T_s$ scale first appear at 150 K, rise
rather slowly as the temperature is lowered, reach about 50\% of their final
value at the superconducting transition and then rise more rapidly.  Finally we
have $T_c$, the temperature of the transition to a true 3D coherent
superconducting state.

To illustrate another phenomenon that seems to follow the $T_s$ scale we have
plotted in the same diagram the amplitude of the 41~meV neutron resonance
peak.\cite{dai00} This peak is loosely called 41~meV resonance but it occurs at
37~meV in underdoped materials. It was originally thought to be associated with
superconductivity since in the optimally doped materials it appeared only below
$T_c$, but as Fig.~3 shows its temperature scale coincides with the spin gap
scale $T_s$ with an onset at 150~K.

%
% Link to magnetism
%
To further explore the various magnetic temperature scales involved in
underdoped cuprates we have replotted in Fig.~4 the pseudogap and the
400~cm$^{-1}$ mode along with two other magnetic quantities, the $^{63}$Cu  NMR
relaxation rate $1/T_1T$ and the $^{63}$Cu Knight shift.\cite{takigawa91} As
pointed out by Homes {\it et al.}, the depth of the pseudogap follows quite
closely the temperature evolution of the NMR Knight shift.  This quantity is
plotted from the work of Takigawa {\it et al.} \cite{takigawa91} in Fig.~4.  On
the other hand we see, as first noted by Hauff {\it et al.}, the 400~cm$^{-1}$
peak tracks the evolution of the $^{63}$Cu relaxation rate.

The idea of two pseudogap temperature scales has been used in the NMR
literature\cite{timusk99} where the lower scale has been designated $T^\ast$
and the higher one $T_0$.  Here we will not follow that practice but will
remain with the more common usage and call the higher scale pseudogap or
$T^\ast$ scale and will rename the lower scale $T_s$ or the spin gap scale.
From Fig.~4 it seems clear that the peak at 400~cm$^{-1}$ follows the $T_s$
scale.

While it appears from Figs.~3 and 4 that the 400~cm$^{-1}$ mode follows the
spin gap there are subtle factors that should be noted.  First the temperature
dependence of the 41~meV mode matches the 400~cm$^{-1}$ peak better than the
spin relaxation $1/T_1T$.  In particular, at $T_c$ there is a rapid rise in
both the 41~meV mode and the 400~cm$^{-1}$ peak, whereas the spin relaxation
rate has no such singularity at $T_c$.

Further support for the two temperature scales is obtained from Zn doping
studies.  It is known from the NMR work of Zheng {\it et al.} that 1\% Zn
doping of YBa$_2$Cu$_4$O$_8$ completely suppresses $T_s$ while leaving the
Knight shift and $T^\ast$ unaffected.\cite{zheng93,zheng96} In accord with
this, Hauff {\it et al.} found that 1.3\% Zn substitution of a
YBa$_2$Cu$_3$O$_{6.56}$ crystal eliminates the 400~cm$^{-1}$ peak (and also
restores the 320 cm$^{-1}$ phonon) while not affecting the amplitude of the
pseudogap.  Thus Zn appears to destroy the spin gap and not affect the
pseudogap.

The effect of Zn on the neutron mode is more complicated.  From what we have
learned from NMR and {\it c}-axis transport we would expect Zn to destroy the
neutron mode.  However, at least in optimally doped materials, Zn drastically
broadens the neutron mode but leaves its spectral weight unchanged up to a
concentration of 1.0\%.\cite{sidis00} This is in accord with measurements of
{\it ab}-plane transport.  In YBa$_2$Cu$_4$O$_8$ even 0.425\% doping has the
effect of dramatically increasing the low temperature scattering rate at
400~cm$^{-1}$ up to the value that it has in the normal state
(1200~cm$^{-1}$).\cite{puchkov96d} Thus it appears that the primary effect of
Zn in the spin gap state is to reduce the carrier life time.

%\subsection{Discussion}

We have found from our analysis of the temperature dependence of the
400~cm$^{-1}$ peak and the vdMM model for the intra-layer plasmon that the {\it
c}-axis data can be understood if there is a current $J_{bl}$ between the
closely spaced layers that starts at 150~K and grows rapidly at $T_c$. What is
the nature of this current?  We can rule out several candidate currents because
they would require an {\it ab}-plane counterpart.  These would include
superconducting fluctuations and collective modes that have to do with sliding
density waves.   The {\it ab}-plane transport shows no diverging conductivity
channels in the 150 to 60~K region in underdoped materials.

The {\it c}-axis dc conductivity matches with the infrared conductivity and the
``semiconducting'' temperature dependence, at least in the 6.60 doping region,
is due to the reduced conductivity as a result of the pseudogap.\cite{homes95}
However within the vdMM model, this overall poor conductivity does not rule out
a highly conducting element in series with a resistive one.  From the good fits
of the model it is clear that there is a high conductivity element between the
closely space bilayers and the interbilayer part is poorly conducting.

%
% Figure 5
%
\begin{figure}[t]
\vspace*{0.5cm}%
\centerline{\includegraphics[width=2.8in]{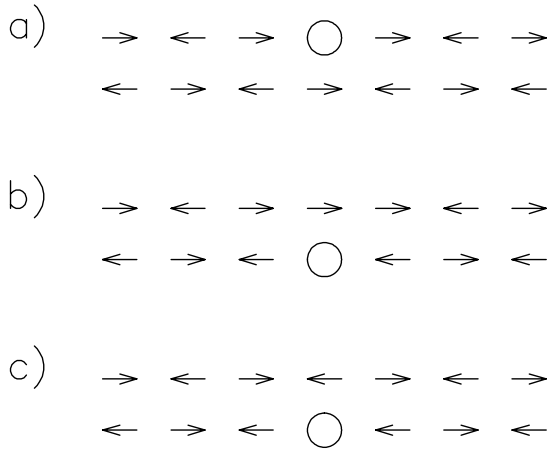}}%
\vspace*{0.5cm}%
%\medskip
\caption{To transfer a hole from layer a) to its neighbor in
antiferromagnetically ordered planes requires an intermediate state of higher
energy b).  A spin flip can restore the ground state c).  We suggest that the
system resonates between states a) and c) forming an odd and even component.
Transitions between these are optically active and may be responsible for the
low lying mode giving rise to the intrabilayer current $J_{bl}$.}%
\label{lumpfig5}
\end{figure}

One possible current path between the bilayers that would be enhanced by long
lived magnetic spin flip excitations is illustrated in Fig.~5. If we assume
that the bilayers are antiferromagnetically correlated then a spin flip is
necessary to move a hole from layer A to layer B. This means that there is an
intermediate state of high energy that has to exist before the
antiferromagnetic order is restored.  The magnetic barrier splits the two
states into an even and odd component and a transition between the two is
optically active. Thus we expect a low lying transition, polarized in the {\it
c} direction that becomes distinct when the barrier becomes well defined, {\it
i.e.} when the various states involved become coherent.  Another current path
is a diagonal second neighbor, one that does not require a spin flip.

The current path shown in Fig.~5 is one of several depending on one's favorite
model for the order that develops in the pseudogap or the spin gap states.  We
are confident that many of these will also contain the ingredients for a
mechanism for a current between the closely spaced layers that yields a {\it
c}-axis current with properties that don't contradict the experiments.

One problem with any model that involves the 41~meV mode directly is the
influence of a static magnetic field.  Neutron scattering carried out in
magnetic fields shows a reduction of the intensity of the resonance.  Dai {\it
et al.} find that a field of 7~T applied in the {\it c} direction has the
effect of reducing the resonance intensity by 20\%, while Kojima {\it et al.}
find, in contrast, that the 400~cm$^{-1}$ peak grows by some 20\% in a
comparable field.  The field {\it reduces} the strength of the neutron
resonance while it {\it enhances} the 400~cm$^{-1}$ peak. Therefore, while the
two show the properties of the state that spans a broad region from some 80~K
above the superconducting transition down to 0~K, the neutron resonance cannot
be a direct cause of the 400~cm$^{-1}$ peak.

%
% Conclusion
%
\section{Conclusion}

We have shown here that at least three temperature scales are needed to
describe the properties of underdoped YBa$_2$Cu$_3$O$_{7-\delta}$.  The largest
of these, traditionally termed $T^\ast$, describes the gap that develops in
{\it c}-axis transport at low frequency.  This scale also fits the NMR Knight
shift as a function of temperature\cite{takigawa91,homes93} and the gaps seen
in other spectroscopies such as tunneling and ARPES.\cite{timusk99}  It is a
broad scale with an onset near room temperature.  ARPES has shown that this gap
has {\it d}-wave symmetry and its appearance in {\it c}-axis transport is a
direct result of the matrix elements of the $d_{x^2-y^2}$ wave functions of the
doped holes.\cite{andersen95,millis99,ioffe00}

The second, lower temperature scale, is associated with spin fluctuations --
nuclear spin relaxation, and the neutron resonance and the 400~cm$^{-1}$ peak.
We call this scale $T_s$.  The phenomena associated with this scale become
observable tens of degrees above the superconducting transition and have their
maximum rate of increase at $T_c$. It is a magnetic scale describing
($\pi,\pi$) spin fluctuations.  It is not clear to us why the intraplane
current should develop a narrow Drude-like response when the spin gap appears
but we have suggested one possibility that involves the 41~meV neutron mode.
Others are possible in the presence of a spin density wave, charge order or
dynamic stripes.  It is important to note that this magnetic scale $T_s$ is
intimately connected to superconductivity, the phenomena that follow this scale
increase rapidly in amplitude at $T_c$. Particularly in optimally doped
materials it is not possible to separate the scales $T_c$ and $T_s$.

Finally, the properties described by the lowest scale appear abruptly at $T_c$,
the superconducting transition temperature.  Long range superconductivity as
shown in our {\it c}-axis experiments by the buildup of spectral weight and by
the delta function contribution to the conductivity at the origin follows this
scale. There is possibly a fourth temperature scale very close to $T_c$ where
2D superconducting fluctuations exist.

We find that the we can fit the 400~cm$^{-1}$ peak in the spectra of Homes {\it
et al.}, in agreement with previous work of Munzar {\it et al.},\cite{munzar99}
to a transverse plasmon arising from current fluctuations between the two
planes of the copper-oxygen bilayer.  The nature of these fluctuations is not
clear to us but they are closely associated with the spin gap. The original
suggestion that they are due to 2D Josephson fluctuation in the planes cannot
be reconciled with the absence of such fluctuations in {\it ab}-plane
transport. Also the presence of the peak in the non-superconducting ladder
compound Sr$_{14-x}$Ca$_x$Cu$_{24}$O$_{41}$ is hard to explain within the
Josephson picture.

%
% Acknowledgements
%
We would thank our colleagues Andy Millis, John Berlinsky, Dimitri Basov,
Takashi Imai, Bruce Gaulin, and Dirk van der Marel for valuable discussions. We
owe particular thanks to Dominik Munzar for help with the details of his model
and for many critical comments. The original crystals for this work were
supplied by the group of Doug Bonn, Walter Hardy and Ruixiang Liang at the
University of British Columbia.
%

%
%%%%%%%%%%%%%%%%%%%%%%%%%%%%%%%%%%%%%%%%%%%%%%%%%%%%%%%%%%%%%%%%%%%%%%%
%
% References
%
%\bibliography{lumpref}

%
% The bibliography, references are in Solid State Communication format
%

%
%
%
\end{document}